\documentclass[a4paper,12pt,fleqn]{article}
\usepackage{amsmath,graphicx}

\newcommand{\pni}{\par\noindent}

\begin{document}
\title{On 4-dimensional cosmological models locally embedded 
in a 5-dimensional Ricci-flat space} \author{ A. G. 
Agnese\footnote{Email: agnese@ge.infn.it}\, and M. La 
Camera\footnote{Email: lacamera@ge.infn.it}} \date{}
\maketitle 
\begin{center}
\emph {Dipartimento di Fisica dell'Universit\`a di 
Genova\\Istituto Nazionale di Fisica Nucleare,Sezione di 
Genova\\Via Dodecaneso 33, 16146 Genova, Italy}\\
\end{center} 
\bigskip
\begin{abstract}\pni
We employ a theorem due to Campbell to build some simple 
4-dimen\-sional cosmological models which originate from 
solutions describing  waves propagating along the 
extra-dimension of a 5-dimensional Ricci-flat space. The 
dimensional reduction is performed in the Jordan frame according 
to the induced-matter theory of Wesson. 
\end{abstract} 
\bigskip\bigskip\bigskip \pni
\vspace{1in}\pni
\newpage
\baselineskip = 2\baselineskip

An interesting version of 5-dimensional General Relativity has 
been developed in recent years by Wesson [1,2,3]. The central 
thesis of his induced-matter theory is that the 4D Einstein's 
field equations with matter
\begin{equation}
G_{\alpha\beta} = 8\pi T_{\alpha\beta}
\end{equation}
are a subset of the 5D field equations for vacuum in terms of
the Ricci tensor
\begin{equation}
R_{AB} = 0
\end{equation}
The theory also allows to obtain flat cosmological solutions 
containing the usual 4D perfect fluid energy-momentum tensor 
[4].\pni
Due to its primary significance in the present context, it is 
worth quoting a theorem due to Campbell [5], which states as 
follows:\pni
\textbf{Theorem:} \emph{Any analytic Riemannian space $V_n(s,t)$ 
can be locally embedded in a Ricci-flat Riemannian space 
$V_{n+1}(s+1,t)$ or $V_{n+1}(s,t+1)$.}\pni
This theorem has been recently brought to light by Romero et al. 
[6,7] and employed both in applying Wesson's method [1] and in 
investigating the embedding of lower-dimensional spacetimes .\pni
In this Brief Note we consider the cosmological solution 
describing  waves propagating in the extra-dimension of a 
(4+1)-dimensional Ricci-flat space and, after selecting 
particular modes, we obtain the corresponding (3+1)-dimensional 
cosmologies. The final result will be written in the Jordan 
frame, according to the dimensional reduction prescribed by the 
``induced-matter theory of Wesson''. \pni
We start from the 5D line element
\begin{equation}
ds_{5}^{2} = e^{\omega}\, (dr^2+d\Omega^2) - e^{\nu}\, dt^2 + 
e^{\mu} \, dl^2 
\end{equation}
where $d\Omega^2 = d\vartheta^2 + \sin{\vartheta}^2 d\varphi^2$ 
and $l$ is the extra coordinate. \pni
The metric coefficients $\omega$, $\nu$ and $\mu$ will depend in 
general on both $t$ and $l$, the dependence on $r$ being ruled 
out in the absence of sources. The case when sources are present
and the metric coefficients depend only on $r$, has been treated 
in a more general context in ref. [8]. Moreover we can put, by 
simmetry considerations, $\nu = \mu$. \pni
The relevant Ricci equations are:
\begin{align*}
&3\, \nu' \omega' - 3\, {\omega'}^2 -2\, \nu'' - 6 \omega'' +3\,
\dot{\nu} \dot{\omega} + 2\, \ddot{\nu} = 0 \tag{4a} \\
&3\, \nu'\omega' + 2\, \nu'' + 3\, \dot{\nu} \dot{\omega} - 3\, 
{\dot{\omega}}^2 - 2\, \ddot{\nu} - 6\, \ddot{\omega} = 0 
\tag{4b} \\ 
&\omega'\dot{\nu} + \nu' \dot{\omega} 
-\omega'\dot{\omega} - 2\, \dot{\omega'} = 0 \tag{4c} \\
&3\,{\omega'}^2 + 2\, \omega'' - 3\, \dot{\omega}^2 - 2 
\ddot{\omega} = 0 \tag{4d} 
\end{align*}
\setcounter{equation}{4}  
Here partial derivatives with respect to $t$ and $l$ are 
denoted by an overdot and a prime respectively.\pni 
One can immediately see that equation (4d) admits 3-brane wave 
solutions, propagating along the fifth dimension of
a 5-dimensional bulk, of the form $\omega_{\pm} = \omega (t \pm 
l)$ and $\nu_{\pm} = \nu (t \pm l)$. Denoting by an asterisk  
derivatives with respect to $t \pm l$, equations (4a), (4b) and 
(4c) all become, after substitution:
 \begin{equation}
2\, {\overset{*}\nu}_\pm {\overset{*}\omega}_\pm - 
{{\overset{*}\omega}_\pm}^2 -2\, {\overset{**}\omega}_\pm = 0 
\end{equation}
Therefore, selecting a particular form of $\omega_\pm$,
the other metric coefficients are given by
\begin{equation}
e^{\nu_\pm} = e^{\mu_\pm} = L\, {\overset{*}\omega}_\pm\, 
e^{\frac{\omega_\pm}{2}} 
\end{equation}
where $L$ is a  suitable constant of integration.\pni
Starting from the above solutions, which describe waves 
propagating along the extra dimension of a 5-dimensional 
Ricci-flat spacetime with line element 
\begin{equation}
ds_{5}^2 = e^{\omega_\pm}\, (dr^2+r^2 d\Omega^2) + L\, 
{\overset{*}\omega}_\pm\, e^{\frac{\omega_\pm}{2}}\,(-dt^2+dl^2)
\end{equation}
we build some simple cosmological models in a 4-dimensional 
spacetime with line element
\begin{equation}
ds_{4}^2 = e^{\omega(t)}\, (dr^2+r^2 d\Omega^2) - 
L\,{\dot{\omega}}(t)\, e^{\frac{\omega(t)}{2}} dt^2
\end{equation}
obtained from (7) by a section with a hypersurface at constant 
$l$ chosen, without loss of generality, as $l=0$.\pni
Components of the Einstein tensor in mixed form, derived from the
above metric (8), are
\begin{equation} \begin{split}
&G_{r}^{r} = G_{\vartheta}^{\vartheta} = G_{\varphi}^{\varphi} =
-\, \dfrac{e^{-\,\frac{\omega}{2}}\, ({\dot{\omega}}^2 + 
\ddot{\omega})}{2\, L\, \dot{\omega}}  \\
&G_{t}^{t} = -\, \dfrac{3\, e^{-\,\frac{\omega}{2}}\, 
\dot{\omega}}{4\, L} 
\end{split} \end{equation}
We wish to match the terms in (9) with the components of the 
usual 4D perfect fluid energy-momentum tensor $T_{\alpha\beta} = 
(p+\rho)\,u_\alpha u_\beta + p g_{\alpha\beta}$. In our case the 
pressure and density are given by $T_{r}^{r} =  p$ and 
$T_{t}^{t} = -\, \rho$, and therefore we can simply identify 
$G_{r}^{r}$ with $8\pi p$ and $G_{t}^{t}$ with $-\, 
8\pi\rho$.\pni Of course the choice of the function $\omega (t)$ 
is to a large extent arbitrary so we suggest, to make some 
physically meaningful examples, the following one:
\begin{equation}
\omega (t) = \alpha \ln {\left(1+\dfrac{t}{\alpha L}\right)}
\end{equation}
where $\alpha$ is an assignable constant.\pni
As a consequence, the line element (8) becomes
\begin{equation}
ds_{4}^{2} = \left(1+\dfrac{t}{\alpha L}\right) 
^{\alpha}\,(dr^2+r^2 d\Omega^2) - \left(1+\dfrac{t}{\alpha L}
\right)^{\frac{\alpha}{2}-1}\, dt^2
\end{equation}
and clearly characterizes a conformally flat spacetime when 
$\alpha + 2 =0$. \pni To go further, it is useful to make in (11)
the change of variable 
\begin{equation}
\tau =\begin{cases}
&\dfrac{4\, \alpha \, L}{\alpha + 2}\, 
\left[\left(1+\dfrac{t}{\alpha L}\right)^{\frac{\alpha + 2}{4}} 
-\, 1 \right] \hspace{2.3cm} \text{if $\alpha +2 \neq 0$} \\ &{}
\\ &-\, 2L \ln{(1-\,\dfrac{t}{2L})} \hspace{4.3cm} \text{if 
$\alpha +2 = 0$} \end{cases}\end{equation} 
thus obtaining
\begin{equation}
ds_{4}^{2}= \left(1+\dfrac{(\alpha +2)\, \tau }{4\,\alpha\, L} 
\right)^{\frac{4 \alpha}{\alpha +2}}\, (dr^2+r^2 d\Omega^2) - 
d\tau^2 \hspace{4.6mm} \text{if $\alpha +2\neq 0$}
\end{equation}
and
\begin{equation}
ds_{4}^{2} = e^{\frac{\tau}{L}}\, (dr^2+r^2 d\Omega^2) - 
d\tau^2 \hspace{36.2mm} \text{if $\alpha + 2 = 0$}
\end{equation} 
where in both cases $1/(2L)$ represents the Hubble constant 
$H_0$. Accordingly, pressure and density of the perfect fluid can
be rewritten respectively as
\begin{equation}
8\pi p = \dfrac{\dfrac{2\,(1-\,\alpha )\,H_0^2} {\alpha 
}}{\left(1+ \dfrac{(\alpha +2)\,H_0 \tau}{2\, \alpha }\right)^2} 
\hspace{1.1cm} 8\pi \rho = \dfrac{3 H_0^2}{\left(1+ 
\dfrac{(\alpha +2)\,H_0 \tau}{2\, \alpha }\right)^2} 
\end{equation}
and
\begin{equation}
8\pi p = -\,3 H_0^2 \hspace{3.1cm} 8\pi \rho = 
3 H_0^2 
\end{equation}
It is apparent that the case $\alpha + 2 = 0$ describes a 
de-Sitter Universe with cosmological constant $\Lambda = 
3 H_{0}^{2}$. On the other hand, the
case $\alpha + 2 \neq 0$, provides the equation of state of 
radiation $3 p = \rho$ when $\alpha = 2/3$, and the equation of 
state of matter $p = 0$ when $\alpha = 1$.

\end{document}